\documentclass[conference]{IEEEtran}
\usepackage[dvips]{graphicx}
\usepackage[cmex10]{amsmath}
\usepackage{cite}
\usepackage{multirow}
\usepackage{epsfig}
\usepackage{mathrsfs}
\usepackage{amssymb}
\usepackage{comment}
\usepackage{bm}
\usepackage{array}
\usepackage{amsthm}
\usepackage{blkarray}
\usepackage{fancyhdr}
\usepackage{enumerate}
\usepackage[lined,boxed,commentsnumbered,ruled,linesnumbered]{algorithm2e}
\usepackage{epsf,psfrag}
\usepackage{epsfig}

\renewcommand{\thispagestyle}[1]{} 

\theoremstyle{definition}
\newtheorem{theorem}{Theorem}[section]
\newtheorem{lemma}[theorem]{Lemma}
\newtheorem{corollary}[theorem]{Corollary}
\newtheorem{proposition}[theorem]{Proposition}

\ifCLASSINFOpdf

\else

\fi

\usepackage[tight,footnotesize]{subfigure}

\begin{document}
\pagestyle{fancy}
\IEEEoverridecommandlockouts

\lhead{\textit{Technical Report, Dept. of EEE, Imperial College, London, UK, Jul., 2012.}}
\rhead{} 
%
\title{Identification of Additive Link Metrics: Proof of Selected Theorems}
\author{\IEEEauthorblockN{Liang Ma\IEEEauthorrefmark{2}, Ting He\IEEEauthorrefmark{3}, Kin K. Leung\IEEEauthorrefmark{2}, Ananthram Swami\IEEEauthorrefmark{4}, and Don Towsley\IEEEauthorrefmark{1}}
\IEEEauthorblockA{\IEEEauthorrefmark{2}Imperial College, London, UK. Email: \{l.ma10, kin.leung\}@imperial.ac.uk\\
\IEEEauthorrefmark{3}IBM T. J. Watson Research Center, Yorktown, NY, USA. Email: the@us.ibm.com\\
\IEEEauthorrefmark{1}University of Massachusetts, Amherst, MA, USA. Email: towsley@cs.umass.edu\\
\IEEEauthorrefmark{4}Army Research Laboratory, Adelphi, MD, USA. Email: ananthram.swami.civ@mail.mil
}
}

\maketitle

\IEEEpeerreviewmaketitle

\section{Introduction}
Selected lemmas, propositions and theorems in \cite{MaNetworkTomography12,MaNetworkTomography12TON} are proved in detail in this report. We first list the lemmas and theorems in Section II and then give the corresponding proofs in Section III. See the original paper \cite{MaNetworkTomography12,MaNetworkTomography12TON} for terms and definitions. Table~\ref{t notion} summarizes all graph theory notations used in this report (following the convention of \cite{GraphTheory2005}).

\begin{table}[tb]
\vspace{-.5em}
\renewcommand{\arraystretch}{1.3}
\caption{Notion in Graph Theory} \label{t notion}
\vspace{-.5em}
\centering
\begin{tabular}{r|m{6.5cm}}
  \hline
  \textbf{Symbol} & \textbf{Meaning} \\
  \hline
  $V$, $L$ & set of nodes/links in the network \\
  \hline
  $|\mathcal{G}|$ & degree of graph $\mathcal{G}$: $|\mathcal{G}|=|V|$ (cardinality of set $V$)\\
  \hline
  $||\mathcal{G}||$ & order of graph $\mathcal{G}$: $||\mathcal{G}||=|L|$\\
  \hline
  $\mathcal{H}$ & interior graph (see Definition 1 in \cite{MaNetworkTomography12})\\
  \hline
  $\mathcal{F}$ & a non-separating cycle (see Definition 2 in \cite{MaNetworkTomography12}) in $\mathcal{G}$\\
  \hline
  $L(v)$ &  set of links incident with node $v$\\
  \hline
  $N_\mathcal{G}(v)$ & set of neighbors of node $v$ in graph $\mathcal{G}$\\
  \hline
  $V(\mathcal{G})$, $L(\mathcal{G})$ & set of nodes/links in graph $\mathcal{G}$\\
  \hline
  $V(L)$ & set of end-points incident with all the links in link set $L$\\
  \hline
  $E(\mathcal{G})$ & set of exterior links (see Definition 1) in graph $\mathcal{G}$\\
  \hline
  $\mathcal{G}-l$ & $\mathcal{G}-l=(V(\mathcal{G}),L(\mathcal{G})\setminus \{l\})$, where $l\in L(\mathcal{G}$) and ``$\setminus$'' is setminus\\
  \hline
  $\mathcal{G}+l$ & $\mathcal{G}+l=(V(\mathcal{G}),L(\mathcal{G})\cup \{l\})$, where $V(\{l\})\subset V(\mathcal{G})$\\
  \hline
  $\mathcal{G}-v$ & $\mathcal{G}-v=(V(\mathcal{G})\setminus \{v\},L(\mathcal{G})\setminus L(v))$, where $v\in V(\mathcal{G})$\\
  \hline
  $\mathcal{G}_s+v$ & $\mathcal{G}_s+v=(V(\mathcal{G}_s)\cup \{v\},L(\mathcal{G}_s)\cup L_v)$, where $\mathcal{G}_s\subset \mathcal{G}$, $v\in V(\mathcal{G}\setminus \mathcal{G}_s)$, and all links incident with $v$ and $V(\mathcal{G}_s)$ in $L(\mathcal{G})$ forms link set $L_v$.\\
  \hline
  $\mathcal{G} \setminus \mathcal{G}^{'}$ & delete all nodes in $V\cap V'$ and their incident links in $\mathcal{G}$\\
  \hline
  $\mathcal{G} \cap \mathcal{G}^{'} $ & intersection of graphs: $ \mathcal{G} \cap \mathcal{G}^{'}=(V \cap V^{'}, L \cap L^{'})$\\
  \hline
  $\mathcal{G} \cup \mathcal{G}^{'} $ & union of graphs: $ \mathcal{G} \cup \mathcal{G}^{'}=(V \cup V^{'}, L \cup L^{'})$\\
  \hline
  $\mathcal{P}=(V,L)$ & simple path $\mathcal{P}$ connecting node $w_0$ and node $w_k$, where $V=\{w_0,\ldots,w_k\}$ and $L=\{w_0w_1,\ldots,w_{k-1}w_k\}$\\
  \hline
  $\mathcal{P}(v_0,v_k)$ & a simple path starting at $v_0$ and terminating at $v_k$ \\
  \hline
  $v_0\underline{a_1\cdots a_k} v_k$ & path $\mathcal{P}(v_0,v_k)$, where $\underline{a_1\cdots a_k}$ is part of the sequenced intermediate nodes or labeled letters along the $v_0$ to $v_k$ direction, is denoted by $v_0\underline{a_1\cdots a_k} v_k$\\
  \hline
  $\mathfrak{S}_\mathcal{P}$ & the sequence of nodes $v_0v_1\ldots v_k$ along path $\mathcal{P}$ \\
  \hline
  $\overset{\circ}{\mathcal{P}}$ & Given path $\mathcal{P}$ with $\mathfrak{S}_\mathcal{P}=v_0v_1\dots v_k$, $\overset{\circ}{\mathcal{P}}$ is a sub-path with $\mathfrak{S}_{\overset{\circ}{\mathcal{P}}}=v_1\dots v_{k-1}$ \\
  \hline
  ${\mathcal{P}\overset{\circ}{v_i}}$ & Given path $\mathcal{P}$ with $\mathfrak{S}_\mathcal{P}=v_0v_1\dots v_k$, ${\mathcal{P}\overset{\circ}{v_i}}$ is a sub-path with $\mathfrak{S}_{\mathcal{P}\overset{\circ}{v_i}}=v_0\dots v_{i-1}$ ($i\in \{0,\cdots,k\}$)\\
  \hline
  $\overset{\circ}{v_i}\mathcal{P}$ & Given path $\mathcal{P}$ with $\mathfrak{S}_\mathcal{P}=v_0v_1\dots v_k$, ${\overset{\circ}{v_i}\mathcal{P}}$ is a sub-path with $\mathfrak{S}_{\overset{\circ}{v_i}\mathcal{P}}=v_{i+1}\dots v_{k}$ ($i\in \{0,\cdots,k\}$)\\
  \hline
  $\mathcal{C}$ & cycle: if $\mathfrak{S}_\mathcal{P}=v_0\cdots v_{k}$ ($k\geq 2$) is a sequence of distinct nodes on path $\mathcal{P}$, then $\mathcal{C}=\mathcal{P}+v_kv_0$ is a \emph{cycle}\\
  \hline
  $W_l$, $W_{\mathcal{P}}$ & link metric on link $l$, sum link metrics on path $\mathcal{P}$\\
  \hline
\end{tabular}
\vspace{-6mm}
\end{table}

\section{Lemmas and Propositions}
Let $\mathcal{H}$ denote the interior graph of graph $\mathcal{G}$, where two monitors ($m_1$ and $m_2$) are employed, and $m^*_1, m^*_2 \in \{m_1,m_2\}$ with $m^*_1\neq m^*_2$. In this report, Conditions \textcircled{\small 1} \normalsize and \textcircled{\small 2} \normalsize refer to the two following conditions.
\begin{description}
  \item[\textcircled{\small 1}]\normalsize $\mathcal{G}-l$ is 2-edge-connected for every interior link $l$ in $\mathcal{H}$;\looseness=-1
  \item[\textcircled{\small 2}]\normalsize $\mathcal{G}+m_1m_2$ is 3-vertex-connected.
\end{description}

\begin{figure}[!thb]
\centering
\includegraphics[width=2.8in]{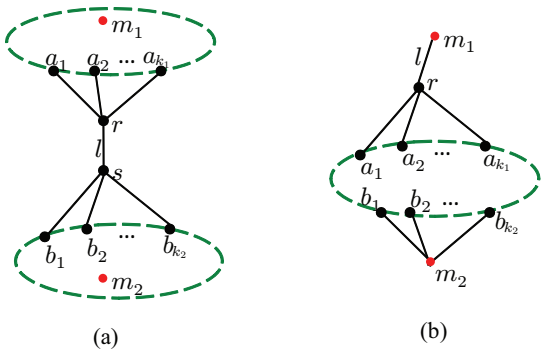}
\caption{Two cases of bridge link $l$: (a) interior bridge, (b) exterior bridge.} \label{fig:Bridge_example}
\end{figure}

\begin{lemma}\label{lemma:BridgeUnidentifiable}
Suppose two monitors are deployed in $\mathcal{G}$ to measure simple paths. If link $l$ is a bridge in $\mathcal{G}$ with one monitor on each side, as illustrated\footnote{In this report, an area with dashed border denotes a sub-graph (the nodes/links within the dashed border are also part of the sub-graph), and a solid line denotes a link/path/cycle.\looseness=-1} in Fig.~\ref{fig:Bridge_example}, then neither $l$ nor its adjacent links are identifiable.
\end{lemma}

\begin{proposition}
\label{lemma:3vertexConnected}
Using two monitors measuring simple paths, if all link metrics in the interior graph of $\mathcal{G}$ are identifiable, then $\mathcal{G}+m_1m_2$ is 3-vertex-connected.
\end{proposition}

\begin{proposition}
\label{Lemma-3-connected}
Using two monitors, the necessary and sufficient condition for $\mathcal{G}+m_1m_2$ being a 3-vertex-connected graph is when 2 nodes are deleted in $\mathcal{G}$, the remaining graph is still connected, \emph{or} every connected component has a monitor.
\end{proposition}

\begin{lemma}
\label{Lemma-twoCycles}
If graph $\mathcal{G}$ satisfies Conditions \textcircled{\small 1} \normalsize and \textcircled{\small 2}\normalsize , then for any interior link $vw$, there exists a non-separating cycle $\mathcal{F}$ with $vw\in L(\mathcal{F})$, a cycle $\mathcal{C}$ with $vw\in L(\mathcal{C})$, a simple path $\mathcal{P}_1$ connecting one monitor with a node on $\mathcal{F}-v-w$, and a simple path $\mathcal{P}_2$ connecting the other monitor with a node on $\mathcal{C}-v-w$ such that
\begin{enumerate}[(a)]
  \item $\mathcal{F}$ and $\mathcal{C}$ have at most one common node other than $v,\: w$ (i.e., $|V(\mathcal{F})\cap V(\mathcal{C})|\leq 3$);
  \item $\mathcal{P}_1$ and $\mathcal{P}_2$ are disjoint, neither going through $v$ nor $w$ (i.e., $\mathcal{P}_1 \cap \mathcal{P}_2=\emptyset$, $v,w\notin V(\mathcal{P}_1)$ and $v,w\notin V(\mathcal{P}_2)$);
  \item  $|V(\mathcal{P}_1)\cap V(\mathcal{F})|=1$, $|V(\mathcal{P}_2)\cap V(\mathcal{C})|=1$.
\end{enumerate}
\end{lemma}

\begin{lemma}
\label{Proposition:RingOneBorderLink}
If $\mathcal{G}$ satisfies Conditions \textcircled{\small 1} \normalsize and \textcircled{\small 2}\normalsize , then
\begin{enumerate}[(a)]
    \item for any non-separating cycle in $\mathcal{G}$, there is at most \emph{one} Case-B link in this non-separating cycle;
    \item for any Case-B link $vw$ in the interior graph of $\mathcal{G}$, there exists a non-separating cycle $\mathcal{F}_{vw}$ with $vw\in L(\mathcal{F}_{vw})$ and $m_1,m_2\notin V(\mathcal{F}_{vw})$. For this non-separating cycle $\mathcal{F}_{vw}$, there exist disjoint simple paths $\mathcal{P}(m^*_1,v)$ and $\mathcal{P}(m^*_2,w)$, each intersecting with $\mathcal{F}_{vw}$ only at the end-point, i.e., $\mathcal{P}(m^*_1,v)\overset{\circ}{v} \cap \mathcal{F}_{vw}=\emptyset$ and $\mathcal{P}(m^*_2,w)\overset{\circ}{w} \cap \mathcal{F}_{vw}=\emptyset$ ($(m^*_1,m^*_2)=(m_1,m_2)$ or $(m_2,m_1)$).
\end{enumerate}
\end{lemma}

\begin{proposition}
\label{Proposition:3EdgeConnectivity}
Given a graph G employing $\kappa$ ($\kappa\geq 3$) monitors, the extended graph $\mathcal{G}_{ex}$ of $\mathcal{G}$ satisfies Conditions \textcircled{\small 1} \normalsize (i.e., $\mathcal{G}_{ex}-l$ is 2-edge-connected for each link $l$ in $\mathcal{G}$) if and only if $\mathcal{G}_{ex}$ is 3-edge-connected.
\end{proposition}

\begin{proposition}
\label{Proposition:3VertexConnectivity}
Given a graph G employing $\kappa$ ($\kappa\geq 3$) monitors, the extended graph $\mathcal{G}_{ex}$ of $\mathcal{G}$ satisfies Conditions \textcircled{\small 2} \normalsize (i.e., $\mathcal{G}_{ex}+m'_1m'_2$ is 3-vertex-connected) if and only if $\mathcal{G}_{ex}$ is 3-vertex-connected.
\end{proposition}


%

\begin{corollary}
\label{corollaryLsLdUnidentifiable}
None of the exterior links (except $m_1m_2$) can be identified with two monitors.
\end{corollary}

\begin{theorem}
\label{theorem:MMP}
For an arbitrary connected network $\mathcal{G}$, Algorithm~MMP generates the \emph{optimal} monitor placement in the sense that: (1) all link metrics in $\mathcal{G}$ are identifiable under this placement, and (2) no placement can identify all link metrics in $\mathcal{G}$ with a smaller number of monitors.
\end{theorem}

\section{Proofs}

\subsection{Proof of Lemma~\ref{lemma:BridgeUnidentifiable}}

The proof follows similar arguments as in the proof of Theorem~3.1 in \cite{MaNetworkTomography12}. Consider the case of Fig.~\ref{fig:Bridge_example}-a. Assume all links except link $l$ and its adjacent links $a_ir$ ($i=1,\ldots,k_1$) and $sb_j$ ($j=1,\ldots,k_2$) are identified. We can then reduce the linear equation associated with any $m_1\to m_2$ path to the form
\begin{align}
W_{a_ir}+W_{sb_j}+W_l=\phi_{ij}, ~~~i=1,\ldots,k_1,\: j=1,\ldots,k_2.
\end{align}
Writing these equations in matrix form and applying the linear transform in the proof of Theorem~III.1 yield a transformed measurement matrix (blank entries are zero):

\scriptsize
\begin{center}
\vspace{-.5em}
$
\left. \begin{aligned}
\mathbf{R}'=&\left.
        \begin{blockarray}{ccccc|cccc|cl}
            \BAmulticolumn{5}{c}{\multirow{1}{*}{$W_{a_1r}\cdots W_{a_{k_1}r}$}}&\BAmulticolumn{4}{c}{\multirow{1}{*}{$W_{sb_1}\cdots W_{sb_{k_2}}$}}& W_l &\\
            \begin{block}{(ccccc|cccc|c)l}
                    1 &   &  & &   & 1 &   &   &  & 1 & \BAmulticolumn{1}{l}{\multirow{5}{*}{ $\left. \begin{aligned} \\ \\ \\ \\ \\ \end{aligned}\right\} \begin{aligned}k_2\\\mbox{rows}\end{aligned}$}}\\
                   1 &   &  & &   &   &  1 &   &  & 1 &\\
                   \vdots & &  &   &   &   &   & \ddots &  & \vdots &\\
                   1 &   & &  &   &   &   &   & 1 & 1 &\\
                   \cline{1-10}
                    & 1  & &  &   &  1 &   &   &  & 1 & \BAmulticolumn{1}{l}{\multirow{5}{*}{ $\left. \begin{aligned} \\ \\ \\ \\ \\ \end{aligned}\right\} \begin{aligned}k_1-1\\\mbox{rows}\end{aligned}$}} \\
                    &   & 1  & &  &  1 &    &   &  & 1 &\\
                    &  &  & \ddots &   & \vdots  &   &   &  & \vdots &\\
                    &   & &  & 1  &  1 &   &   &   & 1 &\\
    \end{block}
  \end{blockarray}\right.\\
  \end{aligned} \right..
$
\vspace{-1.6em}
\end{center}
\normalsize

This matrix corresponds to the maximum set of linearly independent equations involving the unknown variables $W_l$, $(W_{a_ir})_{i=1}^{k_1}$, and $(W_{sb_j})_{j=1}^{k_2}$. Since any subset of $k$ equations contains more than $k$ unknown variables, none of these variables can be identified. Therefore, $l$ and its adjacent links are all unidentifiable.

In the case of Fig.~\ref{fig:Bridge_example}-b, similar argument applies, except that $W_{a_ir}$ is replaced by $W_{ra_i}$ and $W_{sb_j}$ is replaced by $W_{b_j m_2}$. \hfill$\blacksquare$

\subsection{Proof of Proposition \ref{lemma:3vertexConnected}}
\begin{figure}[tb]
\centering
\includegraphics[width=3.4in]{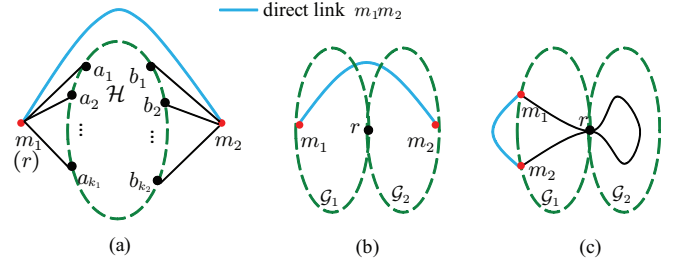}
\caption{Possible scenarios of a cut-vertex $r$.} \label{fig:proof_1vertex}
\end{figure}

\begin{figure}[tb]
\centering
\includegraphics[width=3.1in]{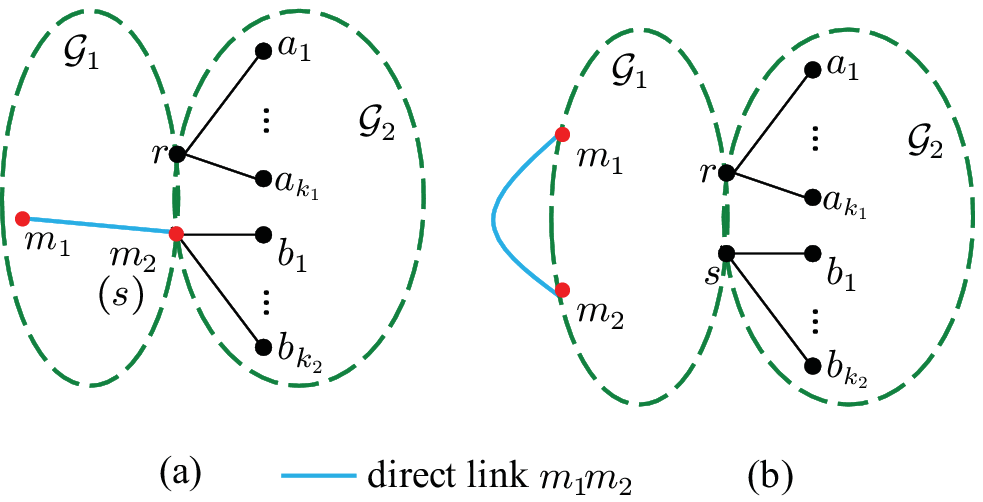}
\caption{Possible scenarios of a 2-vertex cut $\{r,s\}$.} \label{fig:proof_2vertex}
\end{figure}

Suppose that all the interior links of $\mathcal{G}$ are identifiable, and $\mathcal{G}+m_1m_2$ is not 3-vertex-connected. Then the \emph{connectivity}\footnote{The greatest integer $k$ such that $\mathcal{G}$ is $k$-vertex-connected is the \emph{connectivity} of $\mathcal{G}$.} of $\mathcal{G}+m_1m_2$ must be $1$ or $2$.

\emph{1)} Suppose that the connectivity of $\mathcal{G}+m_1m_2$ is $1$. Then it must have a \emph{cut-vertex}, denoted by $r$. There are three possible cases as illustrated in Fig.~\ref{fig:proof_1vertex}. First, if $r\in \{m_1,m_2\}$, as shown in Fig.~\ref{fig:proof_1vertex}-a, then removing $r$ does not disconnect $\mathcal{G}+m_1m_2$ since $\mathcal{H}$ has been assumed to be connected. Thus, $r$ must be in $\mathcal{H}$. Let $\mathcal{G}_1-r$ and $\mathcal{G}_2-r$ denote two of the connected components separated by $r$. If each of $\mathcal{G}_1$ and $\mathcal{G}_2$ contains a monitor, as shown in Fig.~\ref{fig:proof_1vertex}-b, then removing $r$ does not disconnect $\mathcal{G}+m_1m_2$ because link $m_1m_2$ still connects $\mathcal{G}_1$ and $\mathcal{G}_2$; if one of the components, say $\mathcal{G}_2-r$, has no monitor, as shown in Fig.~\ref{fig:proof_1vertex}-c, then any $m_1\rightarrow m_2$ path employing links in $\mathcal{G}_2$ must both enter and leave $\mathcal{G}_2$ through $r$, forming a cycle (which is forbidden). Therefore, the connectivity of $\mathcal{G}+m_1m_2$ must be greater than $1$.

\emph{2)} Suppose that the connectivity of $\mathcal{G}+m_1m_2$ is $2$. Thus, there must be a \emph{2-vertex cut}, denoted by $\{r,s\}$. There are three possibilities. First, if $\{r,s\}=\{m_1,m_2\}$ (Fig.~\ref{fig:proof_1vertex}-a), then removing $\{r,s\}$ does not disconnect $\mathcal{G}+m_1m_2$ because the remaining graph $\mathcal{H}$ is still connected. Second, if $\exists$ one monitor in $\{r,s\}$, say $s=m_2$ as shown in Fig.~\ref{fig:proof_2vertex}-a, then any $m_1\to m_2$ path employing links in $\mathcal{G}_2$ must enter $\mathcal{G}_2$ through $r$ and exit through $m_2$. We can effectively view $r$ and $m_2$ as the new ``monitors'' and their adjacent links $\{ra_i\}_{i=1}^{k_1}$ and $\{m_2 b_j\}_{j=1}^{k_2}$ as the new ``exterior links'' for $\mathcal{G}_2$. Applying Corollary~4.1 in \cite{MaNetworkTomography12} yields that $\{ra_i\}_{i=1}^{k_1}$ and $\{m_2 b_j\}_{j=1}^{k_2}$ are all unidentifiable, contradicting the assumption that all the interior links are identifiable (because $\{ra_i\}_{i=1}^{k_1}$ are interior links). Finally, if both $r$ and $s$ are in $\mathcal{H}$, as shown in Fig.~\ref{fig:proof_2vertex}-b, then any $m_1\to m_2$ path employing links in $\mathcal{G}_2$ must enter/exit $\mathcal{G}_2$ via $r$ and $s$. Note that there must be a component separated by $\{r,s\}$ that has no monitor (e.g., $\mathcal{G}_2$) because otherwise link $m_1m_2$ will keep the components connected after removing $r$ and $s$. For links in $\mathcal{G}_2$, $r$ and $s$ are effectively the new ``monitors'', and thus by Corollary~4.1 in \cite{MaNetworkTomography12} the links $\{ra_i\}_{i=1}^{k_1}$ and $\{sb_j\}_{j=1}^{k_2}$ are unidentifiable. This also contradicts the assumption that all the interior links are identifiable.

Thus, the connectivity of $\mathcal{G}+m_1m_2$ must be greater than 2, i.e., $\mathcal{G}+m_1m_2$ is 3-vertex-connected.
\hfill$\blacksquare$

\subsection{Proof of Proposition~\ref{Lemma-3-connected}}
\emph{Necessary part}.

\emph{1)} If $\mathcal{G}$ is separated by deleting 2 non-monitors, then each component must have a monitor; otherwise, $\mathcal{G}+m_1m_2$ is 2-vertex-connected.

\emph{2)} If one of the deleted nodes is a monitor, then the remaining graph of $\mathcal{G}$ is the same as that of $\mathcal{G}+m_1m_2$ when these two nodes are deleted (because link $m_1m_2$ is deleted). Since $\mathcal{G}+m_1m_2$ is 3-vertex-connected, this remaining graph must be connected.

\emph{3)} If $m_1$ and $m_2$ are deleted, then the remaining graph is the interior graph $\mathcal{H}$, which is connected according to the assumption.

\emph{Sufficient part}.

\emph{1)} If $\mathcal{G}$ is always connected after deleting two nodes, then $\mathcal{G}$ is 3-vertex-connected, so is $\mathcal{G}+m_1m_2$.

\emph{2)} If $\mathcal{G}$ is separated into two connected components after deleting two nodes, and each component has a monitor, then adding link $m_1m_2$ will connect these components again. Therefore, $\mathcal{G}+m_1m_2$ is 3-vertex-connected.
\hfill$\blacksquare$

\subsection{Proof of Lemma \ref{Lemma-twoCycles}}


\emph{1)} In $\mathcal{G}$, we first prove the existence of two cycles $\mathcal{C}_1$ and $\mathcal{C}_2$ with $vw\in L(\mathcal{C}_1)$, $vw\in L(\mathcal{C}_2)$ and $|V(\mathcal{C}_1)\cap V(\mathcal{C}_2)|=2\ or\ 3$.

(1.i). \emph{Existence of $\mathcal{C}_1$.} For $vw\in L(\mathcal{H})$, $\exists$ an H-path\footnote{$\mathcal{P}$ ($||\mathcal{P}||\geq 1$) is an \emph{H-path} of graph $\mathcal{H}$ if $\mathcal{P}$ meets $\mathcal{H}$ exactly in its end-points.} $\mathcal{P}_{vw}$ from $v$ to $w$ in a 2-vertex-connected graph, according to Proposition 3.1.3 \cite{GraphTheory2005} ($\mathcal{G}$ is a 2-vertex-connected graph, since $\mathcal{G}+m_1m_2$ is 3-vertex-connected). Thus, a cycle $\mathcal{C}_1=\mathcal{P}_{vw}+vw$ is formed.

(1.ii). \emph{Existence of $\mathcal{C}_2$.} Suppose $\mathcal{C}_2$ does not exist in $\mathcal{G}$. Then to connect $v$ and $w$, besides using link $vw$, $\mathcal{C}_1-vw$ is the only alternative way. Hence, each link in $\mathcal{C}_1-vw$ must be a bridge in $\mathcal{G}-vw$, contradicting Condition \textcircled{\small 1}\normalsize. Therefore, there exist $\mathcal{C}_1$ with $vw\in L(\mathcal{C}_1)$ and $\mathcal{C}_2$ with $vw\in L(\mathcal{C}_2)$ ($\mathcal{C}_1\neq \mathcal{C}_2$).

(1.iii). \emph{By contradiction, we prove it is impossible that $\mathcal{C}_1$ and $\mathcal{C}_2$ must have more than one common link, i.e., $L(\mathcal{C}_1)\cap L(\mathcal{C}_2)=\{vw\}$.} Suppose $\mathcal{C}_2$ must share some unavoidable\footnote{In this paper, \emph{unavoidable} nodes/links are the common nodes/links between $\mathcal{C}_1$ and $\mathcal{C}_2$ for all possible selections of $\mathcal{C}_1$ and $\mathcal{C}_2$ in $\mathcal{G}$.} common links with $\mathcal{C}_1\setminus \{vw\}$. Let $rs$ be one of these common links (this is the special case that $\mathcal{C}_1$ and $\mathcal{C}_2$ have more than three common nodes). Then if $vw$ is deleted, all possible paths connecting $v$ and $w$ must traverse link $rs$. In this case, $rs$ becomes a bridge, contradicting Condition \textcircled{\small 1}\normalsize.
\begin{figure}[tb]
\centering
\includegraphics[width=1.8in]{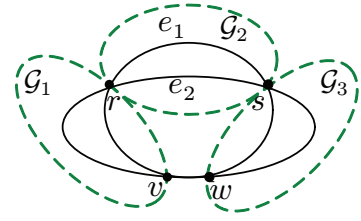}
\caption{Two cycles with four common nodes.} \label{Fig:LemmaTwoCycles}
\end{figure}

(1.iv). \emph{By contradiction, we prove it is impossible that $\mathcal{C}_1$ and $\mathcal{C}_2$ must have more than three common nodes no matter how $\mathcal{C}_1$ and $\mathcal{C}_2$ are selected, i.e., $|V(\mathcal{C}_1)\cap V(\mathcal{C}_2)|=2\ or\ 3$.}

We have $\min(|V(\mathcal{C}_1)\cap V(\mathcal{C}_2)|)=2$, since $vw\in L(\mathcal{C}_1)$ and $vw\in L(\mathcal{C}_2)$. Suppose there are always four common nodes between $\mathcal{C}_1$ and $\mathcal{C}_2$ no matter what strategy is used to select the two cycles. Let $r,s \in V(\mathcal{C}_1)\cap V(\mathcal{C}_2)\setminus \{v,w\}$ (shown in Fig. \ref{Fig:LemmaTwoCycles}) denote the two unavoidable common nodes. Without utilizing link $vw$, to connect $v$ and $w$, all paths leaving from $v$ and terminating at $w$ must first traverse $r$ and then traverse $s$. Therefore, $\mathcal{G}$ can be reformed as sub-graphs $\mathcal{G}_1$, $\mathcal{G}_2$, $\mathcal{G}_3$ (shown in Fig. \ref{Fig:LemmaTwoCycles}) with $V(\mathcal{G}_1\cap \mathcal{G}_2)=r$, $V(\mathcal{G}_2\cap \mathcal{G}_3)=s$, and $\mathcal{G}=\mathcal{G}_1\cup \mathcal{G}_2\cup \mathcal{G}_3+vw$. It has been proved in (1.iii) that $\mathcal{C}_1$ and $\mathcal{C}_2$ cannot have unavoidable common links apart from $vw$, thus each of $\mathcal{G}_1$, $\mathcal{G}_2$, $\mathcal{G}_3$ has at least three nodes. Since at most two sub-graphs, say $\mathcal{G}_1$ and $\mathcal{G}_2$, can contain monitors, nodes in $\mathcal{G}_3-s-w$ are disconnected to monitors, contradicting Proposition~\ref{Lemma-3-connected}. Then obviously, $\mathcal{C}_1$ and $\mathcal{C}_2$ cannot have more than four unavoidable common nodes. Therefore, with the flexibility of selecting all possible $\mathcal{C}_1$ and $\mathcal{C}_2$, it is impossible that $\mathcal{C}_1$ and $\mathcal{C}_2$ must have more than three common nodes.

\emph{2)} With all possible selections, now we prove $\mathcal{C}_1$ and $\mathcal{C}_2$ can be selected to ensure that one of them, say $\mathcal{C}_1$, is a non-separating cycle and the properties of $vw\in L(\mathcal{C}_1)$, $vw\in L(\mathcal{C}_2)$ and $|V(\mathcal{C}_1)\cap V(\mathcal{C}_2)|=2\ or\ 3$ can still be preserved.

\begin{figure}[tb]
\centering
\includegraphics[width=3.1in]{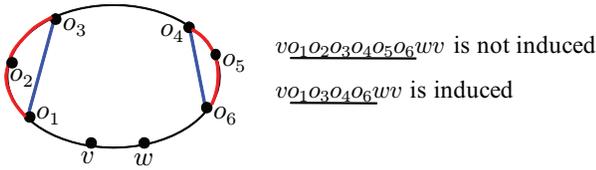}
\caption{Replacement operation to obtain an induced cycle.} \label{Fig:InducedCycle}
\end{figure}

(2.i). \emph{We prove some links on $\mathcal{C}_1$ with $|V(\mathcal{C}_1)\cap V(\mathcal{C}_2)|=2\ or\ 3$ can be replaced to ensure the resulting $\mathcal{C}_1$ is an induced graph.} If $xy\in L(\mathcal{G})$ with $x,y\in V(\mathcal{C}_1)$ and $xy \notin L(\mathcal{C}_1)$, then use $xy$ to replace $\mathcal{P}_{\mathcal{C}_1}(x,y)$ recursively, i.e., $\mathcal{C}_1=\mathcal{C}_1\setminus \overset{\circ}{\mathcal{P}}_{\mathcal{C}_1}(x,y)+xy$, until no such $xy$ exists, where $\mathcal{P}_{\mathcal{C}_1}(x,y)$ is the path from $x$ to $y$ in $\mathcal{C}_1$ with $vw \notin L(\mathcal{P}_{\mathcal{C}_1}(x,y))$. Finally, $\mathcal{C}_1$ is an induced cycle. Note for each replacement operation, link $xy$ does not belong to $\mathcal{C}_2$ since $\mathcal{C}_1$ and $\mathcal{C}_2$ only have 2 or 3 common nodes (two of these two common nodes are $v$ and $w$). Therefore, this replacement operation on $\mathcal{C}_1$ does not affect the number of common nodes between $\mathcal{C}_1$ and $\mathcal{C}_2$. This process is shown is Fig.~\ref{Fig:InducedCycle}, where all red segments are replaced by the blue segments.

\begin{figure}[tb]
\centering
\includegraphics[width=3.1in]{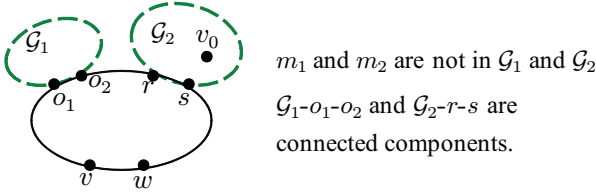}
\caption{Counter example - a non-separating cycle containing $vw$ cannot be found.} \label{Fig:RingExistence}
\end{figure}
(2.ii). \emph{We prove that the induced graph $\mathcal{C}_1$ can be selected to ensure $\mathcal{C}_1$ is a non-separating cycle.} Suppose not. Then for all possible cycles containing $vw$ and satisfying (a) in Lemma \ref{Lemma-twoCycles}, none of them are non-separating cycles. In this case, $\exists$ sub-graph $\mathcal{G}'$ ($m_1,m_2 \notin \mathcal{G}'$) within $\mathcal{G}$ such that when some (not all) nodes in $\mathcal{G}'$ are used to construct $vw$-contained cycles (note that at least 2 nodes in $\mathcal{G}'$ are used), there exist some nodes that are disconnected to monitors in set $A$ ($A\subset V(\mathcal{G}')$) no matter which of these constructed cycles is deleted. This case (shown in Fig. \ref{Fig:RingExistence}) only happens when nodes in $A$ must traverse two nodes in $\mathcal{G}'$, say node $v_0$ in $\mathcal{G}_2$ must use $r$ and $s$, to connect to monitors. Therefore, $v_0$ in $\mathcal{G}_2$ cannot connect to monitors when $r$ and $s$ are deleted, contradicting Proposition~\ref{Lemma-3-connected}. This special case is displayed in Fig. \ref{Fig:RingExistence}, where $\mathcal{G}'=\mathcal{G}_1\cup \mathcal{G}_2$ and $A=V(\mathcal{G}_1\cup \mathcal{G}_2)\setminus \{o_1, o_2, r, s\}$.

\emph{3)} Since cycles can be selected to ensure that $\mathcal{C}_1$ is a non-separating cycle; therefore, $\mathcal{C}_1$ is written as $\mathcal{F}$ instead in the sequel. Since $\mathcal{G}$ is connected, there exist simple path $\mathcal{P}_1$ connecting one monitor and a node on $\mathcal{F}-v-w$, and simple path $\mathcal{P}_2$ connecting the other monitor and a node on $\mathcal{C}_2-v-w$ in $\mathcal{G}$. Note $\mathcal{P}_1$ (or $\mathcal{P}_2$) can be a single node without containing any links. Properties of $\mathcal{F}$ and $\mathcal{C}_2$ are proved in \emph{1)} and \emph{2)}. Now we prove $\mathcal{P}_1$ and $\mathcal{P}_2$ with $\mathcal{P}_1 \cap \mathcal{P}_2=\emptyset$ can be found.

\begin{figure}[tb]
\centering
\includegraphics[width=3.5in]{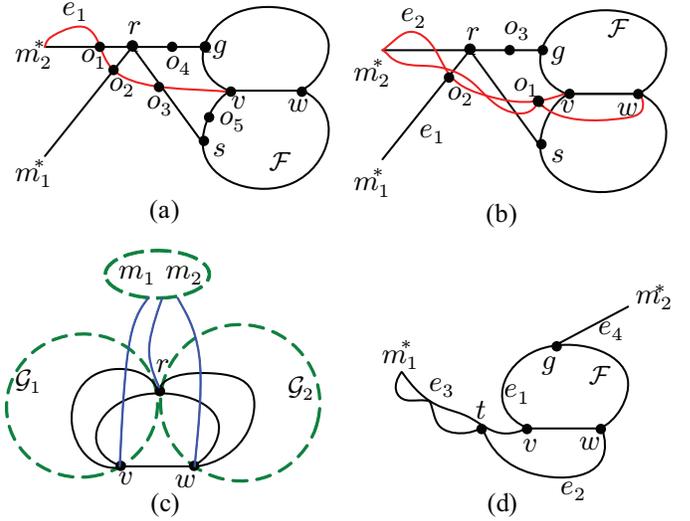}
\caption{Construction of two cycles and two paths.} \label{Fig:LemmaTwoPathsConstruction}
\end{figure}

(3.i). If all $\mathcal{P}_1$ ($m^*_1\in V(\mathcal{P}_1)$) must traverse $m^*_2$, then $m^*_2$ is a cut-vertex in $\mathcal{G}$, contradicting Proposition~\ref{Lemma-3-connected}. Similarly, $m^*_1$ is not an unavoidable node on $\mathcal{P}_2$ ($m^*_2\in V(\mathcal{P}_2)$).

(3.ii). \emph{With the aid of one end-point of $vw$, we prove that $\mathcal{C}_2$ and $\mathcal{P}_2$ can be reselected when $\mathcal{P}_1$ and $\mathcal{P}_2$ have one unavoidable node for the given $\mathcal{F}$ and $\mathcal{C}_2$.} Let $\mathcal{F}=v\underline{s}w+vw$. For any $\mathcal{P}_1$ and $\mathcal{P}_2$, if they must have a common node, say $r$ ($r\notin V(\mathcal{F} \cup \mathcal{C}_2)$, see Fig. \ref{Fig:LemmaTwoPathsConstruction}-a), then $r$ cannot be a cut-vertex, because $\mathcal{G}$ is 2-vertex-connected. Therefore, there must be another path employing $v$ or $w$ (one and only one of $v$ and $w$), say $m^*_2\underline{o_1\cdots o_5v}g$ ($r\notin V( m^*_2\underline{o_1\cdots o_5v}g)$), to connect $m^*_2$ and $g$. $m^*_2\underline{o_1\cdots o_5v}g$ might have common nodes ($o_1,\cdots, o_5$) with other paths. However, if $o_4$ or $o_5$ is the common node, then $\mathcal{P}_1$ and $\mathcal{P}_2$ do not need to traverse $r$ to connect the two cycles. Thus, $o_4$ and $o_5$ are not common nodes. Therefore, $\mathcal{C}_2$ can be reselected, i.e., $\mathcal{C}_2=v\underline{o_1ro_4g}w+vw$ with $\mathcal{P}_2=m^*_2\underline{e_1}o_1$ and $\mathcal{P}_1=m^*_1\underline{r}s$. Note $o_1\underline{ro_4}g$ is impossible to have unavoidable common nodes with $\mathcal{F}$; otherwise, $m^*_2$ cannot connect to $g$ when $v$ and $w$ are deleted. Moreover, $\mathcal{P}_1$ might have common nodes, say $o_2$ or $o_3$, with newly selected $\mathcal{C}_2$. This, however, does not affect the property that $|V(\mathcal{F})\cap V(\mathcal{C}_2)|=2\ or\ 3$, because $\overset{\circ}{v}\underline{o_3o_2o_1ro_4}g$ does not have common nodes with $\mathcal{F}$. Thus, the reselected $\mathcal{C}_2$ will not add any new common nodes between $\mathcal{F}$ and $\mathcal{C}_2$.

(3.iii). \emph{With the aid of all end-points of $vw$, we prove that $\mathcal{C}_2$, $\mathcal{P}_1$ and $\mathcal{P}_2$ can be reselected when $\mathcal{P}_1$ and $\mathcal{P}_2$ have one unavoidable node $r$} ($r\notin V(\mathcal{F} \cup \mathcal{C}_2)$) \emph{for the given $\mathcal{F}$ and $\mathcal{C}_2$.} Let $\mathcal{F}=v\underline{g}w+vw$ (see Fig. \ref{Fig:LemmaTwoPathsConstruction}-b). Suppose $m^*_2$ can make use of both $v$ and $w$, say $m^*_2\underline{o_1}v$ and $m^*_2\underline{o_1}w$, to connect nodes on $\overset{\circ}{v}\underline{g}\overset{\circ}{w}$. We have $r,o_3\notin V(m^*_2\underline{o_1}v\cup m^*_2\underline{o_1}w)$, since $m^*_1$ and $m^*_2$ must traverse $r$ to connect to nodes on $\overset{\circ}{v}\underline{g}\overset{\circ}{w}$ when $v$ and $w$ are not used. In addition, $m^*_2\underline{o_1}v$ and $m^*_2\underline{o_1}w$ do not have common nodes with $\mathcal{F}-v-w$. Therefore, utilizing $m^*_2\underline{o_1}v$ and $m^*_2\underline{o_1}w$, $\mathcal{C}_2$ and $\mathcal{P}_2$ can be reselected, i.e., if $\overset{\circ}{m^*_2}\underline{o_1}v$ and $\overset{\circ}{m^*_2}\underline{o_1}w$ have common nodes, say $o_1$, then $\mathcal{C}_2=v\underline{o_1}w+vw$ with $\mathcal{P}_2=m^*_2\underline{e_2}o_1$; otherwise, $\mathcal{C}_2=v\underline{e_2m^*_2}w+vw$ with $\mathcal{P}_2=\{m^*_2\}$. In the case that $\mathcal{P}_2=m^*_2\underline{e_2}o_1$, if $m^*_2\underline{e_2}o_1$ has a common node with $m^*_1\underline{e_1}r$, say $o_2$, then based on previous operations, we can reselect $\mathcal{P}_1$ and $\mathcal{P}_2$ with $\mathcal{P}_1\cap \mathcal{P}_2=\emptyset$, i.e., $\mathcal{P}_1=m^*_2\underline{ro_3}g$ and $\mathcal{P}_2=m^*_1\underline{e_1o_2}o_1$. While for the unchanged $\mathcal{F}$, it does not have any common nodes with the reselected $\mathcal{C}_2$. Although $\mathcal{P}_1$ might have common nodes with the reselected $\mathcal{C}_2$, the property of $|V(\mathcal{F})\cap V(\mathcal{C}_2)|=2\ or\ 3$ is undamaged.

(3.iv). \emph{We prove the common node between $\mathcal{P}_1$ and $\mathcal{P}_2$ cannot be the common node between $\mathcal{F}$ and $\mathcal{C}_2$.} According to (3.ii)--(3.iii), $\exists\ \mathcal{P}_1$, $\mathcal{P}_2$, and $\mathcal{C}_2$ such that $\mathcal{P}_1$ and $\mathcal{P}_2$ do not have common node $r$ ($r\notin V(\mathcal{F} \cup \mathcal{C}_2)$). However, if $r$ is the common node between $\mathcal{F}$ and $\mathcal{C}_2$ (see Fig. \ref{Fig:LemmaTwoPathsConstruction}-c), we can also prove it is impossible. In this case, as Fig. \ref{Fig:LemmaTwoPathsConstruction}-c shows, $V(\mathcal{G}_1)\cap V(\mathcal{G}_2)=\{r\}$ with $m_1,m_2\notin V(\mathcal{G}_1)$ and $m_1,m_2\notin V(\mathcal{G}_2)$ since $m_1$ and $m_2$ must use $r$ or $v$ or $w$ to connect nodes on $\mathcal{F}-v-w-r$ and $\mathcal{C}_2-v-w-r$. For the two cycles, we have $|\mathcal{G}_1-r-v|\geq 1$ and $|\mathcal{G}_2-r-w|\geq 1$ (since $vw$ is the only common link between $\mathcal{F}$ and $\mathcal{C}_2$); therefore, nodes in $\mathcal{G}_1$ ($\mathcal{G}_2$) without monitors are separated when $r$ and $v$ ($w$) are deleted, contradicting Proposition~\ref{Lemma-3-connected}.

Based on (3.i)--(3.iv), therefore, $\exists\ \mathcal{P}_1$ and $\mathcal{P}_2$ without common nodes, i.e., $\mathcal{P}_1 \cap \mathcal{P}_2=\emptyset$.

\emph{4)} \emph{Now we prove that $v$ and $w$ are not unavoidable nodes on $\mathcal{P}_1$ and $\mathcal{P}_2$, i.e., $v,w\notin V(\mathcal{P}_1)$ and $v,w\notin V(\mathcal{P}_2)$.} We first consider $\mathcal{P}_1$. In $\mathcal{G}-m^*_2$, if $\mathcal{P}_1$ must traverse an end-point of $vw$, say $v$, to connect $m^*_1$ and a node on $\mathcal{F}-v-w$, then nodes on $\mathcal{F}-v-w$ are disconnected to $m^*_1$ when $v$ and $m^*_2$ are deleted, contradicting Proposition~\ref{Lemma-3-connected}. Thus, it is impossible that $\mathcal{P}_1$ must traverse an end-point of $vw$. However, if $\mathcal{P}_1$ cannot avoid traversing one of $v$ and $w$ to connect $m^*_1$ and $\mathcal{F}-v-w$, then two paths can be constructed. Let $v\underline{e_1g}w+vw$ be $\mathcal{F}$ (see Fig. \ref{Fig:LemmaTwoPathsConstruction}-d). The constructed two paths, connecting $m^*_1$ and $g$, are $m^*_1\underline{e_3tve_1}g$ and $m^*_1\underline{e_3te_2w}g$ with $m^*_1\underline{e_3t}v \cap \overset{\circ}{v}\underline{e_1g}\overset{\circ}{w}=\emptyset$ and $m^*_1\underline{e_3te_2}w \cap \overset{\circ}{v}\underline{e_1g}\overset{\circ}{w}=\emptyset$ (if they have intersections, $\mathcal{P}_1$ does not have to traverse $v$ or $w$ to connect to a node on $\overset{\circ}{v}\underline{e_1g}\overset{\circ}{w}$). According to Proposition~\ref{Lemma-3-connected}, $g$ must have a connection to $m^*_2$, $m^*_2\underline{e_4}g$, with $m^*_2\underline{e_4}g \cap m^*_1\underline{e_3}t=\emptyset$ (if $m^*_2\underline{e_4}g \cap m^*_1\underline{e_3}t\neq\emptyset$, then $\mathcal{P}_1$ does not have to traverse $v$ or $w$ to connect to a node on $\overset{\circ}{v}\underline{e_1g}\overset{\circ}{w}$). Therefore, $\mathcal{C}_2$ can be chosen as $\mathcal{C}_2=v\underline{te_2}w+vw$ with $\mathcal{P}_2=m^*_1\underline{e_3}t$ and $\mathcal{P}_1=m^*_2\underline{e_4}g$ (if $\overset{\circ}{m^*_1}\underline{e_3t}v$ and $\overset{\circ}{m^*_1}\underline{e_3te_2}w$ do not have common nodes, then $\mathcal{C}_2=v\underline{e_3m^*_1e_2}w+vw$ with $\mathcal{P}_2=\{m^*_1\}$). These two cycles and paths enable $vw$ to be a cross-link identifiable via the method proposed in Section~5.2.1 of \cite{MaNetworkTomography12}. When considering $\mathcal{P}_2$, the same argument ($vw$ is identifiable via the method proposed in Section~5.2.1 of \cite{MaNetworkTomography12}) applies. Therefore, a cross-link can find $\mathcal{F}$, $\mathcal{C}_2$, $\mathcal{P}_1$ and $\mathcal{P}_2$ with all the properties in Lemma \ref{Lemma-twoCycles}.

\emph{5)} With the properties of (a) and (b) in Lemma~\ref{Lemma-twoCycles}. Let $\mathcal{P}$ be a path starting at one monitor and terminating at a node in $\mathcal{C}-v-w$ with $vw \in L(\mathcal{C})$ and $|V(\mathcal{P})\cap V(\mathcal{C}-v-w)|>1$. In this case, simply use the first common node between $\mathcal{P}$ and $\mathcal{C}$ as the termination of $\mathcal{P}$. Thus, $\mathcal{F}$, $\mathcal{C}_2$, $\mathcal{P}_1$ and $\mathcal{P}_2$ with properties (a), (b) and (c) in Lemma~\ref{Lemma-twoCycles} can be found.
\hfill$\blacksquare$

\subsection{Proof of Lemma \ref{Proposition:RingOneBorderLink}-(a)}

\begin{figure}[tb]
\centering
\includegraphics[width=3.4in]{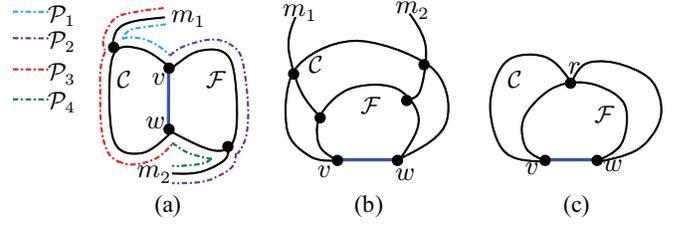}
\caption{Interior link $vw$ is (a) Case-A link, (b) Case-B-1 link, (c) Case-B-2 link. } \label{Fig:Link3Cases}
\end{figure}

Using the method to calculate cross-link (Section~5.2.1 of \cite{MaNetworkTomography12}), all Case-A links in $\mathcal{H}$ can be identified. While for Case-B links, they can be further categorized into two classes: (i) Case-B-1, $V(\mathcal{F} \cap \mathcal{C}_2)=\{v,w\}$ and all $\mathcal{P}_1$ must have a common node with $\mathcal{C}_2$, and (ii) Case-B-2, $V(\mathcal{F} \cap \mathcal{C}_2)=\{v,w,r\}$, where $r$ is another unavoidable common node. Fig.~\ref{Fig:Link3Cases} illustrates three types of interior links.

Let $vw$ be a Case-B link in $\mathcal{H}$ and $vw\in L(\mathcal{F})$. All other links on $\mathcal{F}$ can use the same non-separating cycle when constructing cycles and paths specified by Lemma~\ref{Lemma-twoCycles}, because nodes on $\mathcal{C}'_2$ of other links cannot be disconnected to monitors when $\mathcal{F}$ is deleted.
\begin{figure}[tb]
\centering
\includegraphics[width=3.5in]{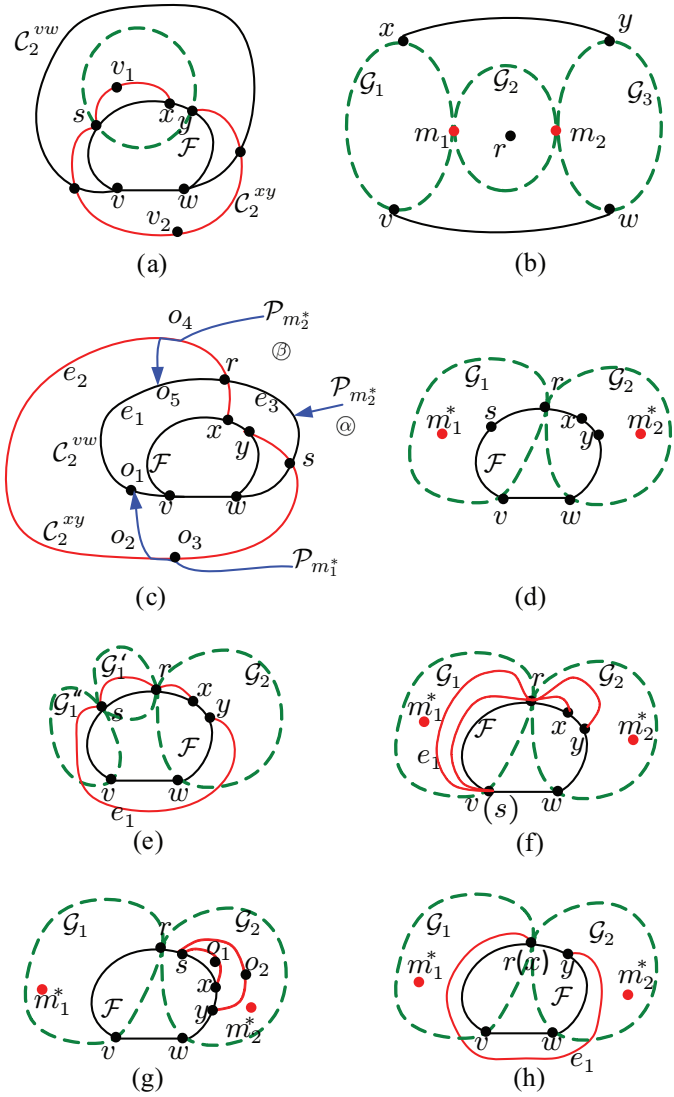}
\caption{Case-B links $vw$ and $xy$ cannot be in the same non-separating cycle.} \label{Fig:PropositionOneRingOneBorder}
\end{figure}

\emph{1)} Let $vw$ be a Case-B-1 link.

(1.i). In Fig. \ref{Fig:PropositionOneRingOneBorder}-a, suppose $xy$ is Case-B-2 link on $\mathcal{F}$, then there is a common node $s$ (there is at most one common node apart from $x$ and $y$, proved in Lemma \ref{Lemma-twoCycles}) on $\mathcal{F}$ and\footnote{Let $\mathcal{C}^{l}$ denote a cycle with link $l\in L(\mathcal{C}^{l})$.} $\mathcal{C}^{xy}_2$. Since $\mathcal{F}$ is an induced graph, there must be a node, say $v_1$, on $s\underline{v_1}x$ and a node, say $v_2$, on $s\underline{v_2}y$. Note the meaning of common node $s$ is that without using link $xy$, $s$ is an unavoidable node to connect $x$ and $y$. Thus, $s\underline{v_1}x$ cannot have common nodes with $\mathcal{C}^{vw}_2$; otherwise, $s$ is not an unavoidable node. Meanwhile, if $v_1$ has a path to one monitor in $\mathcal{G}\setminus \mathcal{C}^{vw}_2$, then $x$ has a path to the same monitor in $\mathcal{G}\setminus \mathcal{C}^{vw}_2$ as well, contradicting the assumption that $vw$ is a Case-B-1 link. Hence, for all paths connecting $v_1$ and monitors, they must traverse $s$ or $y$. Thus, when $s$ and $y$ are deleted, $v_1$ is separated from $m_1$ and $m_2$, contradicting Proposition~\ref{Lemma-3-connected}. This conclusion also holds when $xy$ and $s$ have common nodes with $vw$. As the position of $s$ alters, however, the separated node might change. For instance, when $s=w$, $v_2$ is separated from $m_1$ and $m_2$ when $x$ and $w$ are deleted. Therefore, $xy$ cannot be a Case-B-2 link on non-separating cycle $\mathcal{F}$.

(1.ii). Suppose there is another Case-B-1 link $xy$ on $\mathcal{F}$ and both $\mathcal{C}^{vw}_2$ and $\mathcal{C}^{xy}_2$ must traverse $m_1$ and $m_2$. Then graph $\mathcal{G}$ can be reorganized as Fig. \ref{Fig:PropositionOneRingOneBorder}-b, which is composed of sub-graphs $\mathcal{G}_1$, $\mathcal{G}_2$, $\mathcal{G}_3$ and links $vw$, $xy$. There is at least one node, say $r$, in $\mathcal{G}_2-m_1-m_2$, because we have assumed direct link $m_1m_2$ does not exist in $\mathcal{G}$. Thus, the graph is disconnected when $m_1$ and $m_2$ are deleted, contradicting Proposition~\ref{Lemma-3-connected}. Therefore, it is impossible that $\mathcal{C}^{vw}_2$ and $\mathcal{C}^{xy}_2$ must traverse both $m_1$ and $m_2$.

(1.iii). Since $vw$ is a Case-B-1 link, all possible $\mathcal{P}_1$ must intersect $\mathcal{C}^{vw}_2$. Thus, there exist path $\mathcal{P}_{m^*_1}:=\mathcal{P}(m^*_1, v_1)$ and $\mathcal{P}_{m^*_2}:=\mathcal{P}(m^*_2, v_2)$ with $v_1,v_2\in V(\mathcal{C}^{vw}_2)$ ($v_1$ and $v_2$ can be $v$ and $w$) and $\mathcal{P}_{m^*_1}\cap \mathcal{P}_{m^*_2}=\emptyset$ (If $\mathcal{P}_{m^*_1}\cap \mathcal{P}_{m^*_2}\neq\emptyset$, the common node is a cut-vertex). Suppose there is another Case-B-1 link $xy$ on $\mathcal{F}$ (see Fig. \ref{Fig:PropositionOneRingOneBorder}-c). Then the associated $\mathcal{C}^{xy}_2$ ($V(\mathcal{C}^{xy}_2\cap \mathcal{F})=\{x,y\}$) must have two common nodes (since both $vw$ and $xy$ are Case-B-1 links) with $\mathcal{C}^{vw}_2$, say $r$ and $s$ (we have proved that $r$ and $s$ cannot be both monitors in (1.ii)). Since $xy$ is another Case-B-1 link, if $\mathcal{P}_{m^*_1}$ connects to $\overset{\circ}{r}\underline{e_1vw}\overset{\circ}{s}$, it must have common nodes with $\overset{\circ}{r}\underline{e_2}\overset{\circ}{s}$, say the common node is $o_3$ (the number of common nodes maybe greater than one, say both $o_2$ and $o_3$). In addition, we have $o_3\neq r\neq s$, since if $o_3$ must overlap with $r$ or $s$, then it means $v$ cannot connect to monitors when $r$ and $s$ are deleted, which is impossible. In Fig. \ref{Fig:PropositionOneRingOneBorder}-c, let $o_1$ be another node, which can be equal to $v$, on $\mathcal{C}^{vw}_2$. Now we consider the locations of $\mathcal{P}_{m^*_1}$ and $\mathcal{P}_{m^*_2}$. If $\mathcal{P}_{m^*_2}$ ends at $r\underline{e_3}s$ (location \textcircled{\small $\alpha$} \normalsize in Fig. \ref{Fig:PropositionOneRingOneBorder}-c), then $\mathcal{P}_{m^*_1}$ cannot end at $\overset{\circ}{r}\underline{e_1vw}\overset{\circ}{s}$, because $xy$ can select $x\underline{re_3s}y+xy$ as $\mathcal{C}^{xy}_2$, and then path $m^*_1\underline{o_3o_2o_1}v$ connecting $m^*_1$ and $v$ does not intersect with the newly selected $\mathcal{C}^{xy}_2$, resulting $xy$ to be a non-Case-B-1 link, contradicting the assumption that $xy$ is a Case-B-1 link. Therefore, $\mathcal{P}_{m^*_1}$ also ends at ${r}\underline{e_3}{s}$. In this case, however, $v$ is disconnected to monitors when $r$ and $s$ ($r$ and $s$ cannot be both monitors according to (1.ii)) are deleted, contradicting Proposition~\ref{Lemma-3-connected}. Now we change the location of $\mathcal{P}_{m^*_2}$. If no $\mathcal{P}_{m^*_1}$ and $\mathcal{P}_{m^*_2}$ end at $r\underline{e_3}s$, then both $\mathcal{P}_{m^*_1}$ and $\mathcal{P}_{m^*_2}$ (location \textcircled{\small $\beta$} \normalsize in Fig. \ref{Fig:PropositionOneRingOneBorder}-c) end at $\overset{\circ}{r}\underline{e_1vw}\overset{\circ}{s}$. In this case, $\mathcal{C}^{xy}_2$ can be reselected, i.e., $\mathcal{C}^{xy}_2=x\underline{re_3s}y+xy$ with $\mathcal{P}^{xy}_2=m^*_2\underline{o_4}r$ and $\mathcal{P}^{xy}_1=m^*_1\underline{o_3o_2o_1}v$. Thus, $xy$ with $\mathcal{P}^{xy}_1\cap \mathcal{P}^{xy}_2=\emptyset$, which is a cross-link (Section~5.2.1 of \cite{MaNetworkTomography12}), is not a Case-B-1 link, contradicting the assumption of $xy$ being a Case-B-1 link. This conclusion also holds when $y=w$ (or $x=v$). Thus, $\mathcal{F}$ with Case-B-1 link $vw$ cannot have another Case-B-1 link.

\emph{2)} Let $vw$ be a Case-B-2 link. For $vw$, suppose all cycles must traverse $r$, then $\mathcal{G}$ consists of sub-graph $\mathcal{G}_1$, $\mathcal{G}_2$ and link $vw$ (see Fig. \ref{Fig:PropositionOneRingOneBorder}-d). In addition, each of $\mathcal{G}_1$ and $\mathcal{G}_2$ has a monitor in it; otherwise, $\mathcal{G}_1$ ($\mathcal{G}_2$) is separated from monitors when $r$ and $v$ ($w$) are deleted, contradicting Proposition~\ref{Lemma-3-connected}.

(2.i). Suppose $xy\in L(\mathcal{G}_2)$ (see Fig. \ref{Fig:PropositionOneRingOneBorder}-e) is a Case-B-2 link on the same non-separating cycle $\mathcal{F}$, all $\mathcal{C}^{xy}_2$ must traverse a node, say $s$, on $\mathcal{F}$. If $s$ is on $\overset{\circ}{v}\underline{s}\overset{\circ}{r}$, $\mathcal{G}_1$ is further split into two sub-graphs ($\mathcal{G}'_1$ and $\mathcal{G}''_1$), contradicting the claim that $\mathcal{F}-x-y$ and $\mathcal{C}^{xy}_2-x-y$ cannot have two common nodes. Thus, $s$ cannot be on $\overset{\circ}{v}\underline{s}\overset{\circ}{r}$. If $s=v$, then path $x\underline{e_1}y$ is required. Since $vw$ is a Case-B-2 link, $x\underline{e_1}y$ must traverse $r$ as well, resulting that $\mathcal{C}^{xy}_2-xy$ contains a cycle (see Fig. \ref{Fig:PropositionOneRingOneBorder}-f), contradicting the basic requirement in \cite{MaNetworkTomography12}. To avoid employing cycles, $\mathcal{C}^{xy}_2$ must be in $\mathcal{G}_2$ and $s$ must be on $r\underline{xy}w$ (see Fig. \ref{Fig:PropositionOneRingOneBorder}-g). Since $\mathcal{F}$ is induced, there exist nodes on $\overset{\circ}{s}\underline{o_1}\overset{\circ}{x}$ and $\overset{\circ}{s}\underline{o_2}\overset{\circ}{y}$. According to the properties of a non-separating cycle, $o_1$ and $o_2$ have connections to $m^*_2$ in $\mathcal{G}_2 \setminus \mathcal{F}$ (in Lemma~\ref{Lemma-twoCycles}, we have proved the existence of $\mathcal{F}$). Using these connections (without containing $s$, $x$ and $y$), $\mathcal{C}^{xy}_2$ without traversing $s$ can be found, contradicting the assumption that $xy$ is a Case-B-2 link. When $x=r$ or $y=w$ or $xy\in L(\mathcal{G}_1)$, the same argument can be applied. Thus, $xy$ cannot be a Case-B-2 link on non-separating cycle $\mathcal{F}$ containing a Case-B-2 link $vw$.

(2.ii). Suppose $xy\in L(\mathcal{G}_2)$ (see Fig. \ref{Fig:PropositionOneRingOneBorder}-h) is a Case-B-1 link on the same non-separating cycle $\mathcal{F}$, we have $r=x$ or $r=y$, since $\mathcal{F}$ and $\mathcal{C}^{xy}_2$ cannot have common nodes, apart from $x$ and $y$. If $r=x$, there should be path $r\underline{e_1}y$ and $r\underline{e_1}y$ cannot have any links outside $\mathcal{G}_1$ and $\mathcal{G}_2$; therefore, $r\underline{e_1}y\subset \mathcal{G}_2$. In this case, there is a path $\mathcal{P}(m^*_1,v)$ ($r\notin V(\mathcal{P}(m^*_1,v))$. If $r$ must be on $\mathcal{P}(m^*_1,v)$, then $v$ is disconnected to monitors when $r$ and $w$ are deleted.) connecting $m^*_1$ and $v$ without intersecting $r\underline{e_1}y$, contradicting the assumption that $xy$ is a Case-B-1 link. The same conclusion can be obtained when $r=y$. Thus, $xy$ cannot be a Case-B-1 link on non-separating cycle $\mathcal{F}$ containing a Case-B-2 link $vw$..

Therefore, a non-separating cycle with a Case-B link cannot have another Case-B link.
\hfill$\blacksquare$

\subsection{Proof of Lemma \ref{Proposition:RingOneBorderLink}-(b)}
\label{sect:ringExistence}
\begin{figure}[tb]
\centering
\includegraphics[width=2in]{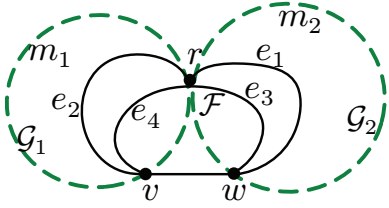}
\caption{Case-B-2 link $vw$ and monitors are not in the same non-separating cycle.} \label{Fig:BorderLinksWithoutMonitors}
\end{figure}

\emph{1) We first prove that for any Case-B link $vw$, there exists a non-separating cycle $\mathcal{F}_{vw}$ with $vw\in L(\mathcal{F}_{vw})$ and $m_1,m_2\notin V(\mathcal{F}_{vw})$.}

If $vw$ is Case-B-1 link, then all paths connecting nodes on $\mathcal{F}_{vw}-v-w$ and monitors must intersect with $\mathcal{C}^{vw}_2$. Therefore, $m_1$ and $m_2$ cannot be on $\mathcal{F}_{vw}$.

If $vw$ is Case-B-2 link and all paths (besides direct link $vw$) connecting $v$ and $w$ must traverse a monitor, say $m_1$, then it means $r=m_1$ in Fig. \ref{Fig:BorderLinksWithoutMonitors}. Thus, $m_2$ is in either $\mathcal{G}_1$ or $\mathcal{G}_2$ (each sub-graph has at least two links; otherwise, the single link becomes a bridge when $vw$ is deleted). Suppose $m_2$ is in $\mathcal{G}_1$, then nodes in $\mathcal{G}_2-r-w$ are disconnected to monitors when $r$ ($r=m_1$) and $v$ are deleted (see Fig. \ref{Fig:BorderLinksWithoutMonitors}), contradicting Proposition~\ref{Lemma-3-connected}. Then obviously, it is impossible that $\mathcal{F}$ must traverse both $m_1$ and $m_2$.

Now suppose either $m_1$ or $m_2$ must be on $\mathcal{F}_{vw}$. Without loss of generality, let $m_1\in V(\mathcal{G}_1)$ and $m_2\in V(\mathcal{G}_2)$ (see Fig. \ref{Fig:BorderLinksWithoutMonitors}). We have proved there exists a non-separating cycle, denoted by $\mathcal{F}$, for any link $vw$ with $vw\in L(\mathcal{F})$ (shown in Fig. \ref{Fig:BorderLinksWithoutMonitors}). Since $\mathcal{F}$ is an induced graph, $r\underline{e_1}w$ cannot be a single link. Thus, there is at least one node on $\overset{\circ}{r}\underline{e_1}\overset{\circ}{w}$. In $\mathcal{G}_2$, when $r\underline{e_3}w$ is deleted, nodes on $\overset{\circ}{r}\underline{e_1}\overset{\circ}{w}$ should be able to connect to monitors (in this case, the monitor can only be $m_2$); otherwise, $v\underline{e_4re_3}w+vw$ is not a non-separating cycle. Therefore, for a non-separating cycle containing Case-B-2 link $vw$, $m_2$ cannot be on $re_3w$. Similarly, $m_1$ cannot be on $re_4v$. Therefore, as long as a non-separating cycle $\mathcal{F}$ containing Case-B-2 link $vw$ can be found (the existence of $\mathcal{F}$ has been proved in Lemma \ref{Lemma-twoCycles}), $\mathcal{F}$ is a non-separating cycle without traversing $m_1$ or $m_2$.

\emph{2) Finally we prove for the non-separating cycle $\mathcal{F}_{vw}$ selected by (b) in Lemma \ref{Proposition:RingOneBorderLink}, there exist simple paths $\mathcal{P}(m_1,v)$ and $\mathcal{P}(m_2,w)$ with $\mathcal{P}(m_1,v) \cap \mathcal{P}(m_2,w)=\emptyset$, $\mathcal{P}(m_1,v)\overset{\circ}{v} \cap \mathcal{F}_{vw}=\emptyset$ and $\mathcal{P}(m_2,w)\overset{\circ}{w} \cap \mathcal{F}_{vw}=\emptyset$.}

\begin{figure}[tb]
\centering
\includegraphics[width=3.4in]{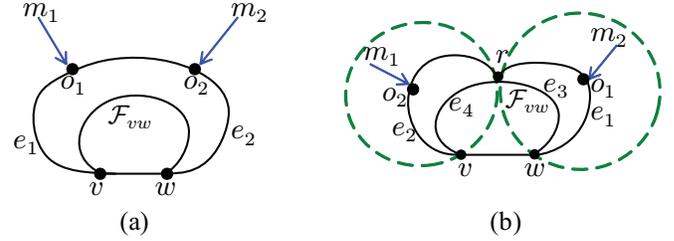}
\caption{Paths construction for Case-B link $vw$.} \label{Fig:PathsConstructionCaseB}
\end{figure}

(2.i). Suppose $vw$ is Case-B-1 link. Then there exist simple paths $\mathcal{P}(m_1,o_1)$ and $\mathcal{P}(m_2,o_2)$, where $o_1,o_2\in V(\mathcal{C}^{vw}_2)$ (see Fig. \ref{Fig:PathsConstructionCaseB}-a). $\mathcal{P}(m_1,o_1)$ and $\mathcal{P}(m_2,o_2)$ do not have unavoidable common node; otherwise, monitors cannot connect to nodes on $\mathcal{C}^{vw}_2$ when the unavoidable common node is deleted. For Case-B link $vw$, we have proved the existence of $\mathcal{F}_{vw}$ with $m_1,m_2\notin V(\mathcal{F}_{vw})$. Therefore, simple paths $\mathcal{P}(m^*_1,v)$ and $\mathcal{P}(m^*_2,w)$ with $\mathcal{P}(m^*_1,v) \cap \mathcal{P}(m^*_2,w)=\emptyset$, $\mathcal{P}(m^*_1,v)\overset{\circ}{v} \cap \mathcal{F}_{vw}=\emptyset$ and $\mathcal{P}(m^*_2,w)\overset{\circ}{w} \cap \mathcal{F}_{vw}=\emptyset$ can be selected, i.e., $\mathcal{P}(m^*_1,v)=m_1\underline{o_1e_1}v$ and $\mathcal{P}(m^*_2,w)=m_2\underline{o_2e_2}w$.

(2.ii). Suppose $vw$ is Case-B-2 link. In Section~\ref{sect:ringExistence}1, we have proved the existence of non-separating cycle $\mathcal{F}_{vw}$ with $vw \in L(\mathcal{F}_{vw})$. In Fig. \ref{Fig:PathsConstructionCaseB}-b, let $F_{vw}=v\underline{e_4re_3}w+vw$. Without loss of generality, let $m_1$ be in $\mathcal{G}_1$ and $m_2$ be in $\mathcal{G}_2$. Then when $re_3w$ is deleted, there must be a path connecting $m_2$ and a node on $\overset{\circ}{r}\underline{e_1}\overset{\circ}{w}$, say $\mathcal{P}(m_2,o_1)$, where $o_1 \in \overset{\circ}{r}\underline{e_1}\overset{\circ}{w}$. Similarly, there exists path $\mathcal{P}(m_1,o_2)$, where $o_2 \in \overset{\circ}{v}\underline{e_2}\overset{\circ}{r}$. Therefore, simple paths $\mathcal{P}(m^*_1,v)$ and $\mathcal{P}(m^*_2,w)$ with $\mathcal{P}(m^*_1,v) \cap \mathcal{P}(m^*_2,w)=\emptyset$, $\mathcal{P}(m^*_1,v)\overset{\circ}{v} \cap \mathcal{F}_{vw}=\emptyset$ and $\mathcal{P}(m^*_2,w)\overset{\circ}{w} \cap \mathcal{F}_{vw}=\emptyset$ can be selected, i.e., $\mathcal{P}(m^*_1,v)=m_1\underline{o_2e_2}v$ and $\mathcal{P}(m^*_2,w)=m_2\underline{o_1e_1}w$.
\hfill$\blacksquare$

\subsection{Proof of Proposition \ref{Proposition:3EdgeConnectivity}}
\emph{Necessary part.} Suppose $\mathcal{G}_{ex}-l$ is 2-edge-connected for all $l$ in $\mathcal{G}$. Consider removing two links in $\mathcal{G}_{ex}$, denoted by $l_1$ and $l_2$.

\emph{1)} If at least one of these links, say $l_1$, is in $L(\mathcal{G})$, then by assumption $\mathcal{G}_{ex}-l_1$ is 2-edge-connected. Thus, $\mathcal{G}_{ex}-l_1-l_2$ is connected.

\emph{2)} Suppose none of these links is in $\mathcal{G}$, i.e., both $l_1$ and $l_2$ are virtual links. Since the virtual monitors $m'_1$ and $m'_2$ each connect to all actual monitors in $\mathcal{G}$, and there are at least $3$ actual monitors, $m'_1$ and $m'_2$ are each connected to $\mathcal{G}$ via at least $3$ virtual links. Therefore, $m'_1$ and $m'_2$ are still connected with $\mathcal{G}$ after $l_1$ and $l_2$ are deleted. Since we have assumed $\mathcal{G}$ to be a connected graph (Section II in \cite{MaNetworkTomography12}), $\mathcal{G}_{ex}-l_1-l_2$ is connected.

We have shown that $\mathcal{G}_{ex}$ remains connected after removing any two links. Therefore, $\mathcal{G}_{ex}$ is 3-edge-connected when $\mathcal{G}_{ex}-l$ ($l \in L(\mathcal{G})$) is 2-edge-connected.

\emph{Sufficient part.}

Suppose $\mathcal{G}_{ex}$ is 3-edge-connected. Then obviously, $\mathcal{G}_{ex}-l$ is 2-edge-connected for each $l \in L(\mathcal{G})$.
\hfill$\blacksquare$

\subsection{Proof of Proposition \ref{Proposition:3VertexConnectivity}}
\emph{Necessary part.}

We prove the necessary part by contradiction. Suppose $\mathcal{G}_{ex}$ is not 3-vertex-connected, but $\mathcal{G}_{ex}+m'_1m'_2$ is 3-vertex-connected, then the connectivity of $\mathcal{G}_{ex}$ must be 2, because removing one link will decrease connectivity by at most 1. Thus, there must exist two nodes, denoted by $v_1$ and $v_2$, whose removal will disconnect $\mathcal{G}_{ex}$. There are 3 possibilities for $v_1$ and $v_2$.

\emph{1)}. If $v_1$, $v_2$ are $m'_1$, $m'_2$, then after their removal, the remaining graph ($\mathcal{G}$) is still connected.

\emph{2)}. If $v_1$ is a virtual monitor ($m'_1$ or $m'_2$) and $v_2$ is a node in $\mathcal{G}$, then $\mathcal{G}_{ex}-v_1-v_2$ being disconnected will imply $\mathcal{G}_{ex}+m'_1m'_2-v_1-v_2$ also being disconnected (as the remaining graphs of $\mathcal{G}_{ex}$ and $\mathcal{G}_{ex}+m'_1m'_2$ are the same), contradicting the assumption that $\mathcal{G}_{ex}+m'_1m'_2$ is 3-vertex-connected.

\begin{figure}[tb]
\centering
\includegraphics[width=3.1in]{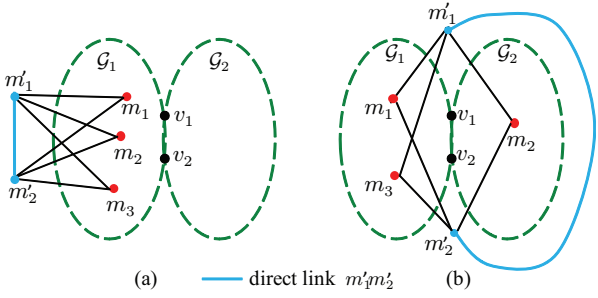}
\caption{$\mathcal{G}_{ex}-v_1-v_2$ is disconnected, where $v_1,v_2\in V(\mathcal{G})$.} \label{Fig:ExtendedGraph3VertexConnectivity}
\end{figure}

\emph{3)}. If $v_1$, $v_2$ are both in $\mathcal{G}$ (can be real monitors), then two cases may occur after removing $v_1$ and $v_2$: (a) $\exists$ a connected component that does not contain any real monitor; (b) each connected component contains at least one real monitor, as illustrated in Fig.~\ref{Fig:ExtendedGraph3VertexConnectivity}.
In the case of Fig.~\ref{Fig:ExtendedGraph3VertexConnectivity}-a, $\mathcal{G}_{ex}+m'_1m'_2-v_1-v_2$ is disconnected as well, contradicting the 3-vertex-connectivity of $\mathcal{G}_{ex}+m'_1m'_2$. In the case of Fig.~\ref{Fig:ExtendedGraph3VertexConnectivity}-b, different components in $\mathcal{G}_{ex}-v_1-v_2$ can still connect via virtual links and virtual monitors, thus contradicting the assumption that $\mathcal{G}_{ex}-v_1-v_2$ is disconnected.

Hence, when $\mathcal{G}_{ex}+m'_1m'_2$ is 3-vertex-connected, the connectivity of $\mathcal{G}_{ex}$ cannot be less than $3$, i.e., $\mathcal{G}_{ex}$ is also 3-vertex-connected.

\emph{Sufficient part.}

If $\mathcal{G}_{ex}$ is 3-vertex-connected, then after adding one link $m'_1m'_2$, $\mathcal{G}_{ex}+m'_1m'_2$ is also 3-vertex-connected.
\hfill$\blacksquare$

\subsection{Proof of Corollary~\ref{corollaryLsLdUnidentifiable}}

Assume that all the interior link metrics are known. We prove Corollary~\ref{corollaryLsLdUnidentifiable} by contradiction. From the proof of Theorem~3.1 in \cite{MaNetworkTomography12}, we see that the transformed measurement matrix $\mathbf{R}'$ gives a maximum set of linearly independent equations (one equation per row) regarding the exterior link metrics $\{W_{m_1a_i}\}^{k_1}_{i=1}$ and $\{W_{b_jm_2}\}^{k_2}_{j=1}$.
Suppose there exists one exterior link metric, say $W_{m_1a_i}$, which is identifiable.
Since each row in $\mathbf{R}'$ only involves two exterior link metrics, it is easy to see that knowing the metric of any exterior link (i.e., $W_{m_1a_i}$ according to the assumption) will allow unique identification of all the other exterior link metrics, contradicting the fact that $\mathbf{R}'$ is rank deficient and hence not able to identify all the metrics. Note the direct link $m_1m_2$ (if exists) is always identifiable. Therefore, none of the exterior links (except $m_1m_2$) can be identified with two monitors.
\hfill$\blacksquare$

\subsection{Proof of Theorem~\ref{theorem:MMP}}
\label{Sect:MMPproof}

It is easy to see from rules~(i)--(iv) (Section~7.2 \cite{MaNetworkTomography12}) that MMP only deploys monitors when needed, and thus no algorithm can achieve identifiability with fewer monitors
Therefore, it suffices to prove the sufficiency of the monitor placement computed by MMP (Algorithm~\ref{Alg:MinimumNumberOfMonitoringNodes}), i.e., all the link metrics can be uniquely identified from measurements between the selected monitors. The idea of our proof is to show that the resulting extended graph $\mathcal{G}_{ex}$ satisfies the condition in Theorem~3.3 \cite{MaNetworkTomography12}.

\begin{algorithm}[tb]
\label{Alg:MinimumNumberOfMonitoringNodes}
\small
\SetKwInOut{Input}{input}\SetKwInOut{Output}{output}
\Input{Connected graph $\mathcal{G}$}
\Output{A subset of nodes in $\mathcal{G}$ as monitors}
choose all the nodes with degree less than $3$ as monitors\; \label{MMP:1}
partition $\mathcal{G}$ into biconnected components $\mathcal{B}_1, \mathcal{B}_2,\ldots$\; \label{MMP:biconnected partition}
\ForEach {biconnected component $\mathcal{B}_i$ with $|\mathcal{B}_i|\geq 3$}
    {partition $\mathcal{B}_i$ into triconnected components $\mathcal{T}_1, \mathcal{T}_2,\ldots$\; \label{MMP:triconnected partition}
    \ForEach {triconnected component $\mathcal{T}_j$ of $\mathcal{B}_i$ with $|\mathcal{T}_j|\geq 3$}
        {
        \If{ $0<s_{\mathcal{T}_j}<3$ \textbf{and} $s_{\mathcal{T}_j}+M_{\mathcal{T}_j}<3$ \label{MMP:2} }
            {
            randomly choose $3-s_{\mathcal{T}_j}-M_{\mathcal{T}_j}$ nodes in $\mathcal{T}_j$ that are neither separation vertices nor monitors as monitors\; \label{MMP:rand1}
            }        \label{MMP:3}
        }
    \If {$0<c_{\mathcal{B}_i}<3$ \textbf{and} $c_{\mathcal{B}_i}+M_{\mathcal{B}_i}<3$ \label{MMP:4}}
        {
        randomly choose $3-c_{\mathcal{B}_i}-M_{\mathcal{B}_i}$ nodes in $\mathcal{B}_i$ that are neither cut-vertices nor monitors as monitors\; \label{MMP:rand2}
        }        \label{MMP:5}
    }
\If {the total number of monitors $K_{\min}<3$ \label{MMP:6}}
    {
    randomly choose $3-K_{\min}$ non-monitor nodes as monitors\; \label{MMP:rand3}
    }           \label{MMP:7}
\caption{Minimum Monitor Placement (MMP)}
\vspace{-.25em}
\end{algorithm}
\normalsize

If $\mathcal{G}$ is already 3-vertex-connected, then $\mathcal{G}_{ex}$ is always 3-vertex-connected as long as there are at least three monitors, no matter how they are placed. This case is handled by lines~\ref{MMP:6}--\ref{MMP:7} in Algorithm~\ref{Alg:MinimumNumberOfMonitoringNodes}. Below we will show that even if $\mathcal{G}$ is not 3-vertex-connected, Algorithm~\ref{Alg:MinimumNumberOfMonitoringNodes} still guarantees that $\mathcal{G}_{ex}$ is 3-vertex-connected.
We prove this statement by showing that after removing any two nodes $u_1,u_2$ in $\mathcal{G}$, each remaining node in $\mathcal{G}-u_1-u_2$ is connected to at least one monitor. There are three possible cases: (1) $u_1,u_2$ belong to the same triconnected component; (2) $u_1,u_2$ belong to different triconnected components within the same biconnected component; (3) $u_1,u_2$ belong to different biconnected components. We now analyze these cases separately.\looseness=-1

(1) Consider deleting two nodes $u_1,u_2$ in a triconnected component $\mathcal{T}$. If $|\mathcal{T}|=2$, then $\mathcal{T}$ is a bridge. According to rules~(i)--(iv) (Section~7.2 \cite{MaNetworkTomography12}), each of the neighboring components of $\mathcal{T}$ must contain at least one monitor other than $u_1$ and $u_2$. Thus, after deleting $u_1$ and $u_2$, each connected component must contain at least one monitor.

\begin{figure}[!thb]
\centering
\includegraphics[width=1.1in]{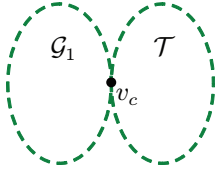}
\vspace{-.5em}
\caption{Possible scenarios for a triconnected component $\mathcal{T}$ with $s_{\mathcal{T}}=1$.} \label{fig:alg1proof_sT1}
\vspace{-.5em}
\end{figure}

\begin{figure}[!thb]
\centering
\includegraphics[width=3.2in]{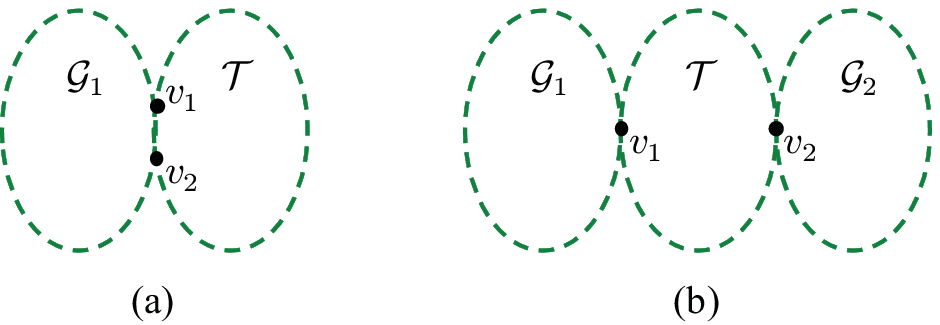}
\vspace{-1em}
\caption{Possible scenarios for a triconnected component $\mathcal{T}$ with $s_{\mathcal{T}}=2$.} \label{fig:alg1proof_sT2}
\end{figure}

\begin{figure}[!thb]
\centering
\includegraphics[width=3.4in]{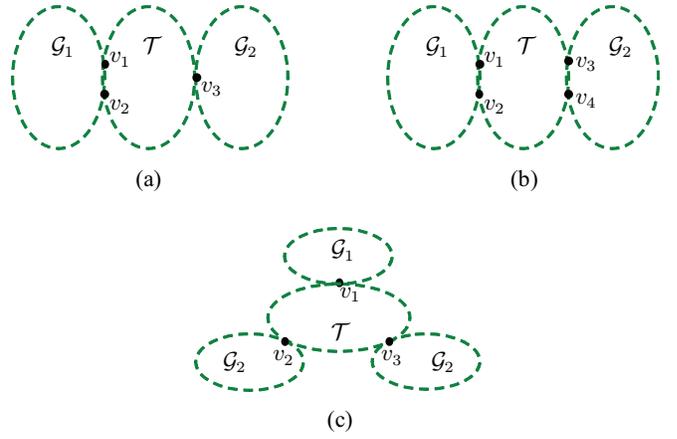}
\vspace{-.5em}
\caption{Possible scenarios for a triconnected component $\mathcal{T}$ with $s_{\mathcal{T}}\geq 3$.} \label{fig:alg1proof_sT3}
\vspace{-.5em}
\end{figure}

Now consider the case of $|\mathcal{T}|\geq 3$. Since $\mathcal{T}-u_1-u_2$ must be connected (due to its triconnectivity), it suffices to show that $\mathcal{T}-u_1-u_2$ contains or is connected to at least one monitor. There are four possibilities depending on the number of separation vertices in $\mathcal{T}$, denoted by $s_{\mathcal{T}}$: \begin{enumerate}[(a)]
\item If $s_{\mathcal{T}}=0$, then $\mathcal{G}$ must be $3$-vertex-connected or a triangle. After removing two vertices, the remaining graph is either connected with at least one monitor, or a degenerate graph of a single node (which is a monitor).\looseness=-1
\item If $s_{\mathcal{T}}=1$, then there exists one cut-vertex $v_c$ in $\mathcal{T}$ (shown\footnote{Note that there may be more than one triconnected subgraph connecting at $v_c$, but we only illustrate the minimum for clarity; the same applies to subsequent illustrations in this proof.} in Fig.~\ref{fig:alg1proof_sT1}). If $v_c\notin\{u_1,u_2\}$, then $\mathcal{T}-u_1-u_2$ is still connected to monitors in neighboring components via $v_c$; if $v_c\in\{u_1,u_2\}$, then $\mathcal{T}-u_1-u_2$ must contain a monitor because two of the non-separation vertices in $\mathcal{T}$ must be monitors by rule~(iv).
\item If $s_{\mathcal{T}}=2$, then there are two possible scenarios (shown in Fig.~\ref{fig:alg1proof_sT2}), i.e., a 2-vertex-cut $\{v_1,v_2\}$, {or} $\mathcal{T}$ has two cut-vertices $v_1$ and $v_2$. If $\{v_1,v_2\} = \{u_1,u_2\}$, then $\mathcal{T}-u_1-u_2$ must contain at least one monitor because $\mathcal{T}$ contains at least one monitor that is not a separation vertex (by rule~(iii)). If $v_2\in\{u_1,u_2\}$ and $v_1\notin\{u_1,u_2\}$ (or the other way), then $\mathcal{T}-u_1-u_2$ is still connected to monitors in neighboring components via $v_1$. Similarly, if $v_1,v_2\notin\{u_1,u_2\}$, then $\mathcal{T}-u_1-u_2$ is still connected to monitors in neighboring components via $v_1$ and $v_2$.
\item If $s_{\mathcal{T}}\geq 3$ (three possible scenarios are shown in Fig.~\ref{fig:alg1proof_sT3}), then Algorithm~\ref{Alg:MinimumNumberOfMonitoringNodes} will not place any monitor in $\mathcal{T}$. Nevertheless, after removing two nodes $u_1$ and $u_2$, at least one separation vertex must remain, and thus $\mathcal{T}-u_1-u_2$ is still connected to monitors in at least one neighboring component via the remaining separation vertex.
\end{enumerate}
The above shows that remaining nodes in $\mathcal{T}$ are still connected to monitors after removing two nodes $u_1$ and $u_2$ within the same triconnected component $\mathcal{T}$.

\vspace{.7em}

(2) Now consider deleting two nodes $u_1,u_2$ in two different triconnected components $\mathcal{T}_i,\mathcal{T}_j$ ($i\neq j$) that are in the same biconnected component $\mathcal{B}$.
Note that the remaining graph $\mathcal{B}'=\mathcal{B}-u_1-u_2$ must be connected, because the only scenario for $\mathcal{B}'$ to be disconnected is when $u_1,u_2$ belong to the same 2-vertex cut of $\mathcal{B}$, which implies that $u_1,u_2$ must belong to the same triconnected component, contradicting our assumption. Similar to Case (1), we have the following four possibilities depending on the number of cut-vertices in $\mathcal{B}$, denoted by $c_{\mathcal{B}}$:\looseness=-1
\begin{enumerate}[(a)]
\item If $c_{\mathcal{B}}=0$, then $\mathcal{G}$ is biconnected ($\mathcal{B}=\mathcal{G}$), and $\mathcal{G}-u_1-u_2$ must contain at least one monitor because the total number of monitors is at least $3$ (see lines~\ref{MMP:6}--\ref{MMP:7}).
\item If $c_{\mathcal{B}}=1$, then rule~(iv) implies that $\mathcal{B}$ ($|\mathcal{B}|\geq 3$) contains two monitors other than the cut-vertex $v_c$. If $v_c\notin\{u_1,u_2\}$, then $\mathcal{B}-u_1-u_2$ is still connected to monitors in neighboring components via $v_c$; if $v_c\in\{u_1,u_2\}$, then $\mathcal{B}-u_1-u_2$ itself contains a monitor.
\item If $c_{\mathcal{B}}=2$, then $\mathcal{B}$ connects to neighboring biconnected components via two cut-vertices $v_1$ and $v_2$. If $v_1,v_2\in\{u_1,u_2\}$, then $\mathcal{B}-u_1-u_2$ itself contains a monitor, as rule~(iii) requires $\mathcal{B}$ to contain a monitor that is not a cut-vertex; if $v_1\notin\{u_1,u_2\}$ (or $v_2\notin\{u_1,u_2\}$), then $\mathcal{B}-u_1-u_2$ is connected to monitors in at least one neighboring component via $v_1$ (or $v_2$).
\item If $c_{\mathcal{B}}\geq 3$, then $\mathcal{B}$ connects to neighboring biconnected components via at least three cut-vertices $v_1$, $v_2$, and $v_3$. There may not be any monitor in $\mathcal{B}$ in this case. However, $\mathcal{B}-u_1-u_2$ must be connected to monitors in at least one neighboring component via the remaining cut-vertex.
\end{enumerate}
The above shows that nodes in $\mathcal{B}-u_1-u_2$ are still connected to monitors after removing nodes $u_1$ and $u_2$ in two different triconnected components but within the same biconnected component $\mathcal{B}$.

Although our argument has focused on the (triconnected or biconnected) component that contains both $u_1$ and $u_2$, it is easy to see that nodes in the neighboring components that share $u_1$ or $u_2$ are also connected to monitors after these two nodes are removed, as this is a special case of the above argument where the number of removed nodes is at most two, and the removed nodes have to be separation vertices.

\emph{Remark:} The above argument assumes that virtual links have been added (see explanations after Definition~5 \cite{MaNetworkTomography12}). However, the construction of virtual links implies that if a node in $\mathcal{B}-u_1-u_2$ will be disconnected from monitors without virtual links, then this node must have only two neighbors, and line~\ref{MMP:1} in Algorithm~\ref{Alg:MinimumNumberOfMonitoringNodes} would have selected this node as a monitor. Thus, our argument holds without added virtual links.

(3) Finally, consider deleting two nodes $u_1$ and $u_2$ from two different biconnected components $\mathcal{B}_1$ and $\mathcal{B}_2$, respectively. Then one of the following scenarios will occur:
\begin{enumerate}[(a)]
\item If neither $u_1$ nor $u_2$ is a cut-vertex\footnote{$u_1$ and $u_2$ cannot be a 2-vertex cut, because otherwise $u_1$ and $u_2$ are in the same triconnected component, contradicting our assumption.}, then $\mathcal{G}-u_1-u_2$ is still connected. Moreover, it contains at least one monitor as there are at least $3$ monitors in $\mathcal{G}$.
\item If $u_1$ is a cut-vertex but $u_2$ is not (or the other way), then we leverage a result from Cases (1)--(2): removing two nodes from a biconnected component (whether in the same triconnected component or not), all remaining nodes are still connected to monitors. Applying this result, we see that removing one node from a biconnected component also does not disconnect any remaining node from monitors, i.e., each connected component in $\mathcal{G}-u_1$ has at least one monitor. Consider the connected component $\mathcal{G}'$ containing $u_2$.
    Removing $u_2$ cannot disconnect $\mathcal{G}'$, as otherwise $u_2$ will have to be in the same biconnected component as $u_1$ or a cut-vertex itself, contradicting our assumptions; furthermore, $\mathcal{G}'$ must contain at least two monitors by rule~(iv). Thus, each connected component in $\mathcal{G}-u_1-u_2$, including $\mathcal{G}'-u_2$, has at least one monitor.
\item If both $u_1$ and $u_2$ are cut-vertices, then each connected component in $\mathcal{G}-u_1-u_2$ must connect to the rest of the graph in the original $\mathcal{G}$ through one or two cut-vertices (i.e., $u_1$ and $u_2$). By rules~(iii)--(iv), each of these connected components contains at least one monitor other than $u_1$ or $u_2$.
\end{enumerate}

In summary, the arguments in Cases (1)--(3) imply that, for any $\mathcal{G}$ (which may not be 3-vertex-connected), Algorithm~\ref{Alg:MinimumNumberOfMonitoringNodes} guarantees that all nodes in $\mathcal{G}-u_1-u_2$ are still connected to monitors after deleting two arbitrary nodes $u_1,u_2$ in $\mathcal{G}$. Since all the monitors are connected through virtual links and virtual monitors in the extended graph $\mathcal{G}_{ex}$, $\mathcal{G}_{ex}-u_1-u_2$ must be connected.
Moreover, if the virtual monitors $m'_1,m'_2$ are deleted from $\mathcal{G}_{ex}$, then the remaining graph (i.e., $\mathcal{G}$) is still connected since $\mathcal{G}$ is assumed to be connected (see Section~2.1 \cite{MaNetworkTomography12}). Finally, if $m'_1$ (or $m'_2$) and $u\in V(\mathcal{G})$ are deleted, then we have shown that all the nodes in $\mathcal{G}-u$ are connected to monitors, and since these monitors are still connected via $m'_2$, the entire graph $\mathcal{G}_{ex}-m'_1-u$ is connected. Therefore, employing Algorithm~\ref{Alg:MinimumNumberOfMonitoringNodes} to place monitors, the resulting extended graph $\mathcal{G}_{ex}$ is always 3-vertex-connected. By Theorem~3.3 in \cite{MaNetworkTomography12}, this implies that the placed monitors can uniquely identify all the link metrics in $\mathcal{G}$.
\hfill$\blacksquare$

\bibliographystyle{IEEEtran}
\bibliography{mybibSimplified}
\end{document}